\documentclass[journal=apchd5,manuscript=article]{achemso}

\usepackage{graphicx}
\usepackage{dcolumn}
\usepackage{amsmath}
\usepackage{amssymb}
\usepackage{bm}

\title{Selection rules for structured light\\in nanooligomers and other nanosystems}

\author{Stephanie Reich}
\affiliation{Freie Universit\"at Berlin, Department of Physics, Arnimallee 14, 14195 Berlin, Germany}
\email{reich@physik.fu-berlin.de}

\author{Niclas S. Mueller}
\affiliation{Freie Universit\"at Berlin, Department of Physics, Arnimallee 14, 14195 Berlin, Germany}

\author{Michal Bubula}
\affiliation{Freie Universit\"at Berlin, Department of Physics, Arnimallee 14, 14195 Berlin, Germany}
\SectionNumbersOff

\date{\today}

\begin{document}
\begin{abstract}
Structured light are custom light fields where the phase, polarization, and intensity vary with position. It has been used for nanotweezers, nanoscale imaging, and quantum information technology, but its role in exciting optical transitions in materials has been little examined so far. Here we use group theory to derive the optical selection rules for nanosystems that get excited by structured light. If the size of the nanostructure is comparable to the light wavelength, it will sample the full beam profile during excitation with profound consequences on optical excitations. Using nanooligomers as model nanosystems, we show that structured light excites optical transitions that are forbidden for linearly polarized or unpolarized light. Such dipole forbidden modes have longer lifetimes and narrower resonances than dipole allowed transitions. We derive symmetry-adapted eigenmodes for nanooligomers containing up to six monomers. Our study includes tables with selection rules for cylindrical vector beams, for beams with orbital angular momentum, and for field retardation along the propagation direction. We discuss multi-photon processes of nonlinear optics in addition to one-photon absorption. Structured light will unlock a broad range of excitations in nanooligomers and other nanostructures that are currently inaccessible to optical studies. 
\end{abstract}

\section{Introduction}

Exciting materials by light is one of the most fundamental ways to study their physical properties. With light we can prepare distinct excited electronic states and follow their evolution over time.  An electromagnetic transition occurs if the superposition between the charge distribution of the initial and final state matches the structure of the exciting field.\cite{Schmiegelow2016,Rochester2001} The transitions are characterized by the multipole structure of the electromagnetic field: Dipole transitions are induced by the oscillating field, quadrupole transitions by the oscillating field gradients, and so on.\cite{Rochester2001} The rates of dipole transitions are by orders of magnitude larger than of the higher-order multipoles and generally dominate the response of materials.\cite{Rivera2016} The scope of our work is, therefore, to examine transitions that get induced by the field amplitude.

The  states that are accessible to dipole excitations are restricted to a set of ``optically active'', ``dipole-allowed'', or ``bright'' transitions that readily interact with unpolarized radiation of moderate intensity.\cite{InuiBook,CardonaBuch,NovotnyBook2012} The  subset of dipole-allowed excitations is identified by optical selection rules. They are derived from the symmetry of the material and the  dipole moment of the photon as its external perturbation.\cite{InuiBook,CardonaBuch,ReichBuch} The standard optical selection rules are based on the fundamental assumption that the electromagnetic field is constant over the characteristic length scale of the material. Since the size of molecules and crystal unit cells is $\lesssim 1\,$nm this is an excellent assumption for visible and infrared photons with a vacuum wavelength $\lambda>400\,$nm.  To increase the number and the type of available excitations, we need to construct situations where the electric field  amplitude varies over the characteristic length scale of the material. One recent proposal was to shrink the wavelength of light so that the field varies more rapidly in space along its propagation direction.\cite{Rivera2016} We will explore another possibility and study selection rules when changing the in-plane spatial distribution of the electric field.

Structured light describes light beams where the phase and polarization profile vary across the beam profile.\cite{Zhan2009,Yao2011} Cylindrical vector beams, for example, are laser beams where the polarization has cylindrical symmetry.\cite{Zhan2009} In radial polarization the electric field points towards the beam center; in azimuthal polarization the electric field is oriented tangentially to the beam.\cite{Zhan2009,NovotnyBook2012} Another form of structured light are beams with a helical phase structure, which means that the beams carry orbital angular momentum.\cite{Yao2011,Allen1992,OAMCollection2017} 
Despite its varying polarization and phase, structured light excites the same  dipole transitions in traditional materials as linearly polarized light. Because the photon field is huge compared to the material system, it only samples the local linear polarization and not the entire polarization profile. Interestingly, this is different for quadrupole transitions that are induced by the more rapidly varying field gradients. Ionic quadrupole transitions experimentally showed a strong dependence on the helical phase structure of the exciting beam.\cite{Schmiegelow2016,Afanasev2018}

Nanotechnology introduced artificial systems with dimensions $1-100\,$nm into physics, materials science, and many other fields. The optical excitations of such nanoscale structures are of particular interest due to their  well-defined mode character and the confinement of the electromagnetic field.\cite{NovotnyBook2012,MaierBuch} With hundreds of nanometers, the size of the structures becomes comparable to the photon wavelength and the quasi-static approximation of constant field no longer applies. Structured light indeed excites optically forbidden or dark modes of nanoscale systems that are inaccessible to unpolarized and linearly polarized light.
\cite{Volpe2009,Parramon2012,Hentschel2013,Yanai2014,Gomez2013,Deng2018,Kerber2017,Kerber2018,Machado2018}  So far, these excitations have been studied in a case-by-case manner using numerical simulations and experiments. Universal, symmetry-derived selection rules beyond the quasi-static dipole approximation remain missing.

In this paper, we present the symmetry-imposed selection rules for optical absorption including retardation and spatial variation in the field. We study nanostructures that get excited by cylindrical vector beams, light with orbital angular momentum, and field retardation. To do so, we first construct the symmetry-derived eigenmodes of  nanoscale oligomers. We consider modes that are induced by the dipole and the quadrupole of the  monomer and discuss the general extension to higher-order electric multipoles. We then derive the selection rules for  dipole-induced absorption and scattering by structured light. We calculate exemplary excitation spectra in nanoplasmonic systems using finite-difference time-domain (FDTD) techniques and discuss the properties of nominally bright and dark modes in the spectra. In addition to linear optics we present the selection rules for non-linear multi-photon processes. We predict second-harmonic generation in centrosymmetric structures when nanooligomers are excited by two photons of $\pm1$ difference in total angular momentum.  Our findings apply to any system as long as the spatial extension of the excited state is a considerable fraction of the photon wavelength and beam focus. To make the paper more accessible, we focus on plasmonic excitations in nanoscale metallic oligomers. Our formalism may be extended to other excitations of interest like plasmon-enhanced optical processes and dielectric nanophotonics.

Metal nanostructures have been studied for their intriguing optical properties as much as their potential photonic application in fields ranging from analytic chemistry and sensing to quantum information technology.\cite{MaierBuch,Halas2011,Tame2013} Light excites localized surface plasmon resonances in metal nanostructures, which are collective oscillations of the metal free electrons.\cite{MaierBuch,NovotnyBook2012} These excitations strongly absorb and scatter photons. They also induce electromagnetic near fields in close vicinity to the metal surface ($<50$\,nm for visible light).  Many applications of plasmonics implicitly or explicitly exploit the near-field excitation. Among the most prominent examples is surface- (or plasmon-) enhanced Raman scattering (SERS), where the plasmonic near field enhances the Raman process by up to ten orders of magnitude.\cite{EtchegoinBuch,Langer2019,LeRu2006,MuellerFaraday2018,Zhu2014}

Plasmonic oligomers are regular arrangements of plasmonic building blocks like  particles, triangles, and discs.\cite{Prodan2003,MaierBuch} They are extremely helpful to understand light-matter interaction in nanosystems, because they allow to construct plasmon eigenstates in a rational way and are straightforward to fabricate.\cite{Prodan2003,Guerrero2012,Zohar2014} In an oligomer the electromagnetic near fields of close-by monomers interact and collective electromagnetic states emerge.\cite{Prodan2003,Hentschel2011,Guerrero2012,Forestiere2013,Zohar2014,Lamowski2018,Pascale2019} The formation of these  states resembles the construction of molecular electronic orbitals from the valence wave functions of the atoms: The oligomer eigenmodes are symmetric and antisymmetric combinations of the optical excitation in the monomers.\cite{Brandl2006,Gomez2010} The bonding configurations have eigenergies below the energy of the monomer excitation; the antibonding configurations are higher in energy. Oligomers are fabricated through the assembly of solution-processed nanoparticles (spheres, cubes, rods, stars etc.) or through the nanofabrication of assemblies of discs, squares, and bars using electron-beam lithography.\cite{Hentschel2011,Hentschel2013,Shafiei2013,Schietinger2009} They typically extend over several 100\,nm and sample the distribution of phase and polarization for visible light.\cite{MuellerFaradayDiscussions2019,Yanai2014,Parramon2012,Kerber2017} Structured light excites dipole-forbidden plasmons as shown for cylindrical vector beams\cite{Gomez2013,Hentschel2013} and light with orbital angular momentum.\cite{Kerber2017,Kerber2018} Recent work on the absorption of light by self-organized nanoparticle layers considered retardation effects and the change of optical selection rules due to the finite wavelength of light.\cite{MuellerACSPhotonics2018,MuellerFaradayDiscussions2019}

\section{Methods}\label{SEC:methods}

We combine the symmetry analysis of plasmonic oligomers, structured beam profiles, field-retardation, and multi-photon processes with simulations of plasmon eigenmodes, optical absorption, and  light scattering. Our symmetry analysis requires straightforward manipulations of group theory: Reducing representations, finding the representations of higher-order multipoles, finding induced representations for a symmetric arrangement of building blocks, and projecting eigenstates. These tools are described in many textbooks on group theory. We recommend Refs.~\citenum{InuiBook, WilsonBook}. For projecting symmetry-adapted eigenstates, we use graphical projection operators, as explained by Reich~\textit{et al.}\cite{ReichBuch}. Two online resources facilitate group theory manipulations like reducing representations, obtaining higher-order moments and so forth: The Bilbao Crystallographic Server\cite{BilbaoCrystI, BilbaoCrystII} and the tables for point groups compiled by Gernot Katzer.\cite{KatzerOnline} For the $D_{2h}$ point group we use $z$ as the basis function for $B_{1u}$, $y$ for $B_{2u}$, and $x$ for $B_{3u}$, which is the convention most commonly found in the group-theory literature. 

\begin{figure}
  \includegraphics[width=7.5cm]{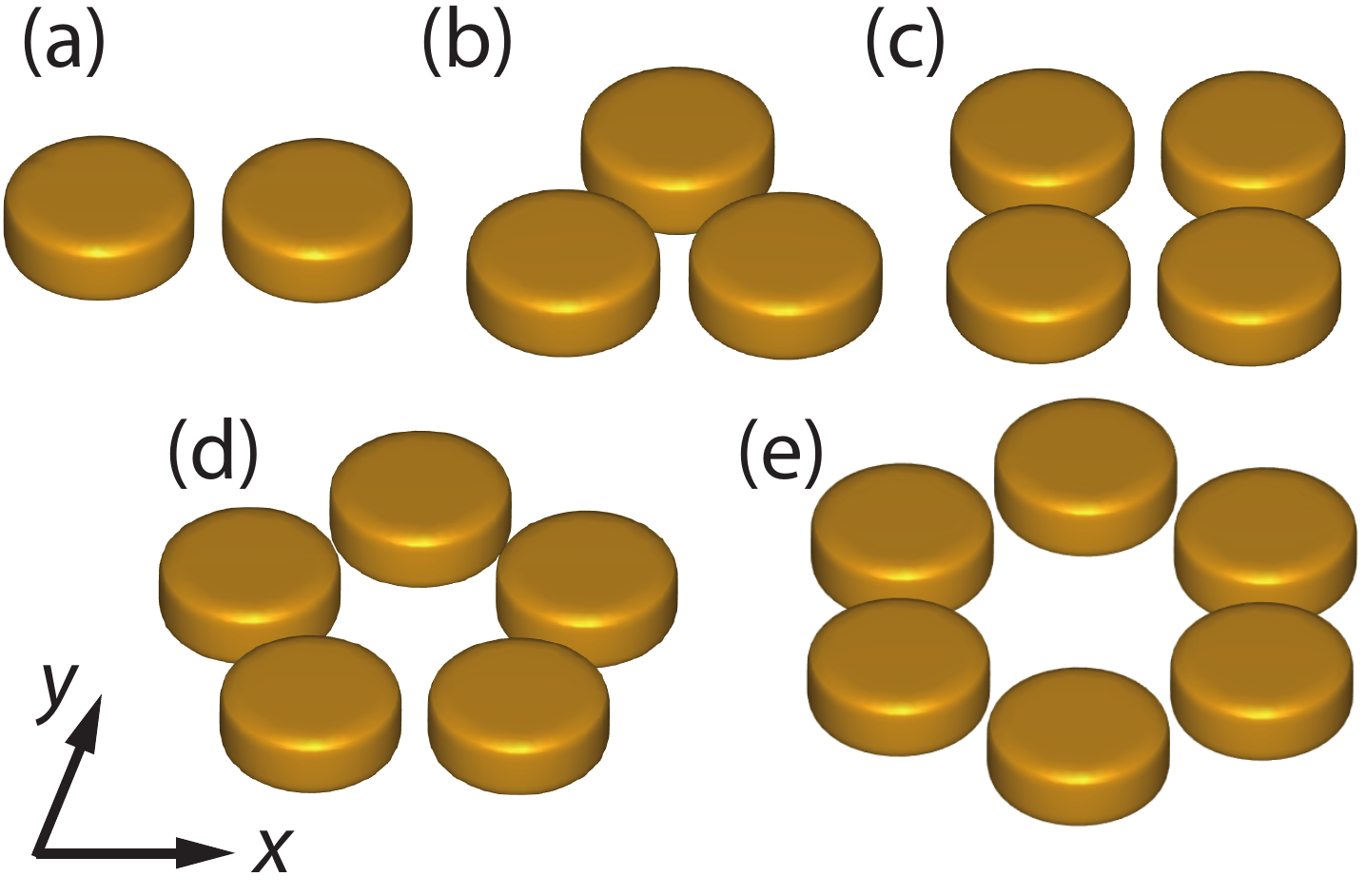}
\caption{Plasmonic oligomers constructed from nanodiscs. (a) Dimer belonging to the $D_{2h}$ point group, (b) trimer ($D_{3h}$), (c) tetramer ($D_{4h}$), (d) pentamer ($D_{5h}$), and (e) hexamer ($D_{6h}$). The geometry of the discs ($d=100\,$nm, $h=40\,$nm) and their separation ($g=20\,$nm) is identical for all oligomers. The arrows indicate the $(x,y)$ coordinate system used throughout the paper except in the section on field retardation.}
\label{FIG:oligomers}
\end{figure}

In addition to the molecular Sch\"onflies notation for point groups of finite systems, we used the formalism that has been developed in connection with line groups of one-dimensional systems.\cite{DamnjanovicBuch,ReichBuch,Damnjanovic1999,Bozovic1985} We briefly introduce the notation for the $D_{nh}$ point groups that are in the focos of our work, see Ref.~\citenum{DamnjanovicBuch} for an extended introduction. The irreducible representations of $D_{nh}$ may be specified by combinig the quantum number $m$ and the parities under horizontal $\sigma_h$ and vertical $\sigma_v$ mirror operation. $m$ can be identified with the $z$ component of the angular momentum along the principle axis of rotation.\cite{DamnjanovicBuch,Damnjanovic1999,ReichBuch} Irreducible representations that are denoted by $A$ in the Sch\"onflies notation have $m=0$ and $B$ $m=n/2$; only $D_{nh}$ groups with even $n$ have $B$ representations. Representations that are denoted by $E_i$ have $m=i$; the subscript gets dropped for $D_{3h}$ ($D_{4h}$) where $E'$ ($E$) has $m=1$. The parity of the mirror operations are either $+1$ or $-1$ for the non-degenerate $A$ and $B$ representations. For the $E$ representations the parity may also be undefined with a character of zero. There is a one-to-one correspondence between the set of $m$ and the two parities and the Sch\"onflies notation that we present in Suppl. Table S1 for the relevant point groups. We give all selection rules and final results in the paper in the Sch\"onflies notation. The line group notation is particularly powerful to analyse optical excitations by beams with orbital angular momentum (OAM). The $m$ quantum number for such a beam corresponds to the combined angular momentum of orbit (i.e., OAM) and spin (polarization). The line group formalism, therefore, greatly facilitates finding the selection rules for OAMs compared to a direct evaluation of the polarization patterns.

We simulated the optical properties for a set of plasmonic oligomers that were constructed from gold nanodiscs, see Fig.~\ref{FIG:oligomers}. The gold discs had a diameter $d=100\,$nm and height $h=40\,$nm. We arranged them in highly symmetric oligomers as shown in Figs.~\ref{FIG:oligomers}(a)-(e) using $g=20\,$nm gaps between adjacent discs. The background dielectric constant $\varepsilon=1.65$ mimicks a dielectric like SiO$_2$ as the oligomer substrate. The simulations use the dielectric function of gold measured by Johnson and Christy.\cite{JohnsonChristy1972} We numerically calculated absorption and light scattering by plasmonic oligomers using the finite-difference time-domain (FDTD) method as implemented in Lumerical. We used a mesh-override region with 2 nm cells to discretize the space around the plasmonic oligomers. For excitation with linearly polarized light, we used a total-field scattered-field plane wave source. For excitation with structured light, i.e. cylindrical vector beams with radial and azimuthal polarization, we used a customized total-field scattered-field source based on a k-space method.\cite{Mansuripur1986} The method is suited to calculate the field distribution near the focus of high numerical-aperture objectives. We implemented cylindrical vector beams with a doughnut radius of $\sim$ 700\,nm at the position of the oligomer. The optical cross sections were recorded with power monitors. The scattering and absorption coefficients were calculated by dividing the cross sections by the area of the oligomer discs $A = n\pi d^2/4$, where $n$ is the number of discs and $d$ their diameter. Plasmon eigenmodes, including their surface charge-density distribution, were obtained with the boundary-elements method, using the eigenmode solver of the MNPBEM Matlab package. \cite{MNPBEM2012} 

To fit plasmon eigenenergies in the absorption spectra we subtracted a background due to the interband transitions of gold. We obtained the background functional form by calculating a slab of gold using identical parameters as for the oligomer simulation.\cite{Bruno2019} The absorption spectra were fit by one (azimuthal, radial polarization) and three (linear) Lorentzian peaks. The scattering spectra were fit with the analytic model by Pinchuk~\textit{et al.}\cite{Pinchuk2004}, which we extended to the case of several plasmon resonances, with amplitudes and frequencies as fitting parameters, see Suppl. Information.

\section{Plasmonic eigenmodes}\label{SEC:gt}

In this section we show how to obtain the symmetry-adapted eigenmodes of a nanooligomer from the excitations of the monomer.\cite{Prodan2003,Guerrero2012,Forestiere2013,Zohar2014} We project the dipole and quadrupole excitations of the discs onto representations of the oligomer using graphical projection operators.\cite{ReichBuch} The approach can be applied to all other multipoles as well. The construction of oligomer eigenmodes within the hybridization model is often restricted to combinations of dipole excitations in the monomers.\cite{Brandl2006,Zohar2014,Haran2018} This assumes that only optically active eigenmodes of the monomer will induce optically active modes of the oligomer, but this is actually not the case.   Dipole-inactive monomer modes like a quadrupole combine in an oligomer into a mode with a finite dipole moment. The collective mode will interact with far-field radiation even if the monomer excitation was dark. The symmetry-adapted eigenvectors are compared to simulated modes obtained by the boundary elements method. 

\subsection{Irreducible representations of plasmons in nanooligomers}

\begin{table*}
\caption{Selected point groups of nanooligomers, plasmonic tips, and colloidal crystals; example structures are given for each point group. The irreducible representations of the dipole and quadrupole moments within each point group are necessary to construct the symmetry-adapted plasmonic or dielectric eigenvectors. ``in-plane" (positive parity under $\sigma_h$) and ``out-of-plane" (negative parity under $\sigma_h$) refer to the $(x,y)$ plane.}
\label{TAB:moments}
\begin{tabular}{llcccccc}\hline\hline
point&example structures&\multicolumn{2}{c}{dipole}&\multicolumn{2}{c}{quadrupole}\\
group&&\multicolumn{2}{c}{representations}&\multicolumn{2}{c}{representations}\\
&&in-plane&out-of-&in-plane&out-of-\\
&&&plane&&plane\\\hline
$D_{2h}$&disc dimer, bowtie&$B_{2u}\oplus B_{3u}$&$B_{1u}$&$2A_{g}\oplus B_{1g}$&$B_{2g}\oplus B_{3g}$\\
&disc chain, dagger\\
$D_{3h}$&trimer&$E'$&$A''_2$&$A_1'\oplus E'$&$E''$\\
$D_{4h}$&tetramer, cross, square&$E_u$&$A_{2u}$&$A_{1g}\oplus B_{1g}\oplus B_{2g}$&$E_g$\\
$D_{5h}$&pentamer&$E_1'$&$A_2''$&$A_1'\oplus E_2'$&$E_1''$\\
$D_{6h}$&hexamer, hexagon, colloidal&$E_{1u}$&$A_{2u}$&$A_{1g}\oplus E_{2g}$&$E_{1g}$\\
&hcp layer, bilayer, crystal\\
$D_{\infty h}$&sphere dimer, tip and image&$E_{1u}/\Pi_u$&$A_{1u}/\Sigma^+_u$&$A_{1g}/\Sigma^+_g \oplus E_{2g}/\Delta_g$&$E_{1g}/\Pi_g$\\
$C_{2v}$&asymmetric disc dimer&$B_1\oplus B_2$&$A_1\oplus $&$2A_1\oplus A_2$&$B_1\oplus B_2$\\
$C_{\infty v}$&tip&$E_1/\Pi$&$A_1/\Sigma^+$&$A_1/\Sigma^+\oplus E_2/\Delta$&$E_1/\Pi $\\\hline\hline
\end{tabular}
\end{table*}

The optical excitations of nanooligomers can be described in a basis of electric multipoles.  The multipoles in the monomer give rise to a set of collective eigenmodes in the oligomer that we will find with the help of the oligomer symmetry. 
We consider an oligomer that is composed of $n$ monomers (nanoparticles, discs, rods) arranged in a symmetric fashion as shown in Fig.~\ref{FIG:oligomers}. Each monomer has many electric multipole excitations that combined will yield the  excitations of the oligomer. In the language of group theory, the representations of the electric multipoles of the monomer induce the symmetry-adapted  eigenmodes of the oligomer. Table~\ref{TAB:moments} lists the point groups for the oligomers in Fig.~\ref{FIG:oligomers} and other nanoplasmonic structures. The table also gives the representation of the dipole and quadrupole moment in each group, which we need to find the representations of the oligomer eigenmodes. To find the eigenmodes, we first set up and reduce the atomic representation $\Gamma_{ar}$, see Table \ref{TAB:plasmon_irreps}.\cite{InuiBook, WilsonBook, CardonaBuch, ReichBuch} It describes the permutation of the monomers under the symmetry operations of the oligomer.\cite{CardonaBuch} The characters of the atomic representation are found by counting the monomers that are left unchanged (= they remain in their original position) by each symmetry operation of the point group. The atomic representation has to be combined with the multipole representation $\Gamma_{mult}$ of the monomer. We  restrict the multipoles to the dipole and quadrupole excitations of the disc, but distinguish between the in-plane and out-of-plane components, see Table \ref{TAB:moments}. The oligomer representation induced by a multipole component $\Gamma_{mult}^{i/o}$ is then given by 
\begin{equation}
\Gamma_{pl}(mult,i/o)=\Gamma_{mult}^{i/o}\otimes \Gamma_{ar},\label{EQ:pl}
\end{equation}
where $i/o$ specifies in-plane and out-of-plane, respectively. Reducing $\Gamma_{pl}$ yields the irreducible representations of the plasmonic eigenmodes. We performed this analysis for the oligomers in Fig.~\ref{FIG:oligomers}, a linear disc trimer, and a nanosphere dimer. The symmetry of the eigenstates that are induced in the oligomers by the dipole and quadrupole excitation of the disc or sphere are given in Table~\ref{TAB:plasmon_irreps}.

Group theory predicts a set of symmetry-adapted eigenmodes that get induced by the monomer multipoles.\cite{InuiBook, WilsonBook}  We project them using graphical projection operators.\cite{ReichBuch} In this method one starts from a graphical representation of the multipole in the monomer. We show the surface charge distribution $+/-$ as red/blue. When applying the symmetry operations of the point group, the starting monomer with its charge distribution is transformed into the other monomers. To project onto a given non-degenerate representation, the charge distribution pattern is multiplied by the character of the representation. A character of $+1$ leaves the pattern unchanged, whereas a character of $-1$ transforms red into blue and vice versa. Summing over all patterns yields an eigenvector of the irreducible representation. The formal treatment and the projection to degenerate representations are discussed in Ref.~\citenum{ReichBuch}.

\subsection{Dipole- and quadrupole-induced eigenmodes}

\begingroup
\begin{table}
\caption{Irreducible representation of plasmonic or dielectric eigenmodes that are induced by the dipole and quadrupole representation of the monomer. The table includes the atomic representation $\Gamma_{ar}$ of selected oligomers; it differentiates between in-plane and out-of-plane moments, see Table~\ref{TAB:moments}.}
\label{TAB:plasmon_irreps}
\begin{tabular}{lc}\hline\hline
&disc dimer $D_{2h}$\\\hline
$\Gamma_{ar}$&$A_g\oplus B_{3u}$\\
dipole, in-plane&$A_g\oplus B_{1g}\oplus B_{2u}\oplus B_{3u}$\\
dipole, out-of-plane&$B_{2g}\oplus B_{1u}$\\
quad., in-plane&$2A_g\oplus B_{1g}\oplus B_{2u}\oplus2 B_{3u}$\\
quad., out-of-plane&$B_{2g}\oplus B_{3g}\oplus A_u\oplus B_{1u}$\\\hline
&linear trimer $D_{2h}$\\\hline
$\Gamma_{ar}$&$2A_g\oplus B_{3u}$\\
dipole, in-plane&$A_g\oplus B_{1g}\oplus 2B_{2u}\oplus 2B_{3u}$\\
dipole, out-of-plane&$B_{2g}\oplus 2B_{1u}$\\
quad., in-plane&$4A_g\oplus 2B_{1g}\oplus B_{2u}\oplus2 B_{3u}$\\
quad., out-of-plane&$2B_{2g}\oplus 2B_{3g}\oplus A_u\oplus B_{1u}$\\\hline
&trimer $D_{3h}$\\\hline
$\Gamma_{ar}$ &$A_1'\oplus E'$\\
dipole, in-plane&$A_1'\oplus A_2'\oplus 2E'$\\
dipole, out-of-plane&$A_2''\oplus E''$\\
quad., in-plane&$2A'_1\oplus A'_2\oplus 3E'$\\
quad., out-of-plane&$A''_1\oplus A''_2\oplus 2 E''$\\\hline
&tetramer $D_{4h}$\\\hline
$\Gamma_{ar}$ &$A_{1g}\oplus B_{2g}\oplus E_u$\\
dipole, in-plane&$A_{1g}\oplus A_{2g}\oplus B_{1g}\oplus B_{2g}\oplus2 E_u$\\
dipole, out-of-plane&$E_g\oplus A_{2u}\oplus B_{1u}$\\
quad., in-plane&$2A_{1g}\oplus A_{2g}\oplus B_{1g}\oplus 2B_{2g}\oplus 3E_u$\\
quad., out-of-plane&$A_{1u}\oplus A_{2u}\oplus B_{1u}\oplus B_{2u}\oplus 2E_g$\\\hline
&pentamer $D_{5h}$\\\hline
$\Gamma_{ar}$ &$A'_1\oplus E'_1\oplus E'_2$\\
dipole, in-plane&$A'_1\oplus A'_2\oplus 2E'_1\oplus 2E'_2$\\
dipole, out-of-plane&$A''_2\oplus E''_1\oplus E''_2$\\
quad., in-plane&$2A'_1\oplus A'_2\oplus 3E'_1\oplus 3E'_2$\\
quad., out-of-plane&$A''_1\oplus A''_2\oplus 2E''_1\oplus 2 E''_2$\\\hline
&hexamer $D_{6h}$\\\hline
$\Gamma_{ar}$ &$A_{1g}\oplus E_{2g}\oplus B_{2u}\oplus E_{1u}$\\
dipole, in-plane&$A_{1g}\oplus A_{2g}\oplus 2E_{2g}\oplus B_{1u}\oplus B_{2u}\oplus 2E_{1u}$\\
dipole, out-of-plane&$B_{1g}\oplus E_{1g}\oplus A_{2u}\oplus E_{2u}$\\
quad., in-plane&$2A_{1g}\oplus A_{2g}\oplus 3E_{2g}\oplus B_{1u}\oplus 2B_{2u}\oplus 3E_{1u}$\\
quad., out-of-plane&$B_{1g}\oplus B_{2g}\oplus 2 E_{1g}\oplus A_{1u}\oplus A_{2u}\oplus 2E_{2u}$\\\hline
&nanosphere dimer, gap mode $D_{\infty}$\\\hline
$\Gamma_{ar}$&$A_{1u}\oplus A_{1g}$\\
dipole, in-plane&$E_{1g}\oplus E_{1u}$\\
dipole, out-of-plane&$A_{1g}\oplus A_{1u}$\\
quad., in-plane&$A_{1g}\oplus E_{2g}\oplus A_{1u}\oplus E_{2u}$\\
quad., out-of-plane&$E_{1g}\oplus E_{1u}$\\\hline\hline
\end{tabular}
\end{table}
\endgroup

\begin{figure}
  \includegraphics[width=8.5cm]{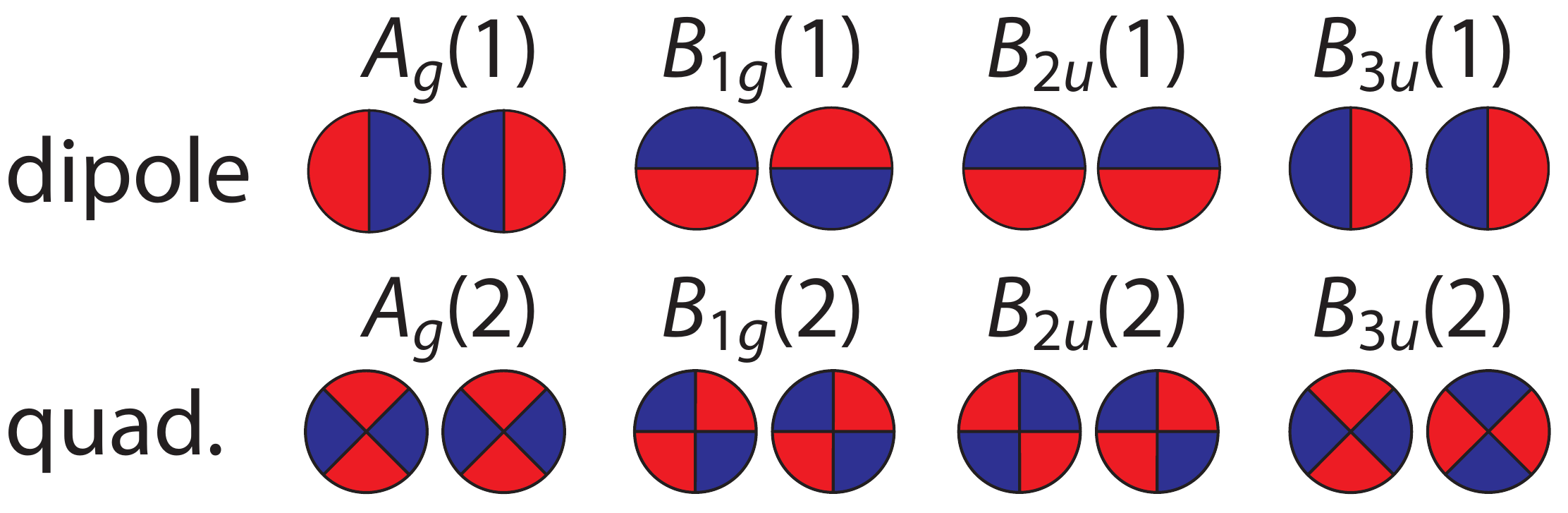}
\caption{Symmetry-adapted eigenmodes of a dimer that are induced by the in-plane monomer dipole (top) and quadrupole (bottom) excitation. The colors represent the sign of the surface charge distribution: red for positive and blue for negative charges. The real charge distribution will differ, because eigenmodes of identical symmetry are allowed to mix.}
\label{FIG:dimer_em}
\end{figure}

\begin{figure}
  \includegraphics[width=8.5cm]{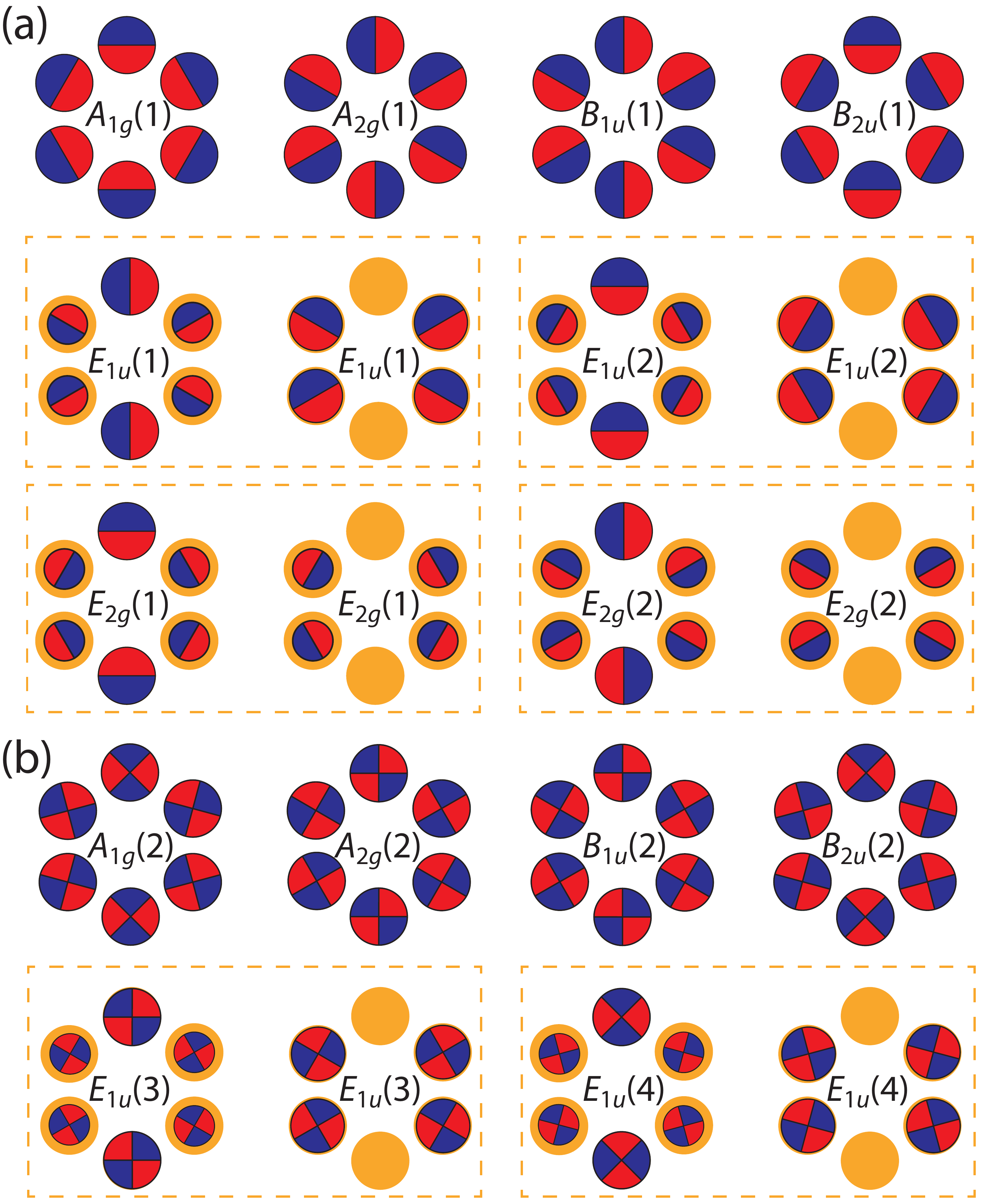}
\caption{Symmetry-adapted eigenvectors of a hexamer that are induced by the in-plane component of the (a) dipole and (b) quadrupole moment in a nanodisc. The multipoles are represented through positive (red) and negative (blue) surface charges. The area of the pattern indicates the relative amplitude of the multipole in each monomer. The yellow circles represent the disc monomer; a full yellow circle means that the eigenmode has zero amplitude at this point and no visible yellow represents maximum amplitude.}
\label{FIG:hexamer_em}
\end{figure}

To demonstrate the eigenmode analysis and the use of projection operators for nanooligomers we consider a disc dimer and hexamer, Fig.~\ref{FIG:oligomers}. The dimer belongs to the $D_{2h}$ point group and has an atomic representation, see Table~\ref{TAB:plasmon_irreps},
\begin{equation*}
    \Gamma_{ar}=A_g\oplus B_{3u}.
\end{equation*}
The in-plane dipole transforms according to $B_{2u}\oplus B_{3u}$ within $D_{2h}$, Table~\ref{TAB:moments}. We obtain a total of four in-plane dipole-induced oligomer representations $\Gamma_{pl}(dip, i)=A_g\oplus B_{1g}\oplus B_{2u}\oplus B_{3u}$. When projecting the monomer dipole onto these irreducible representations, we find the well-known set of four non-degenerate dipolar eigenmodes, see Fig.~\ref{FIG:dimer_em}: $x$-polarized anti-bonding $A_{g}(1)$, $y$-polarized bonding $B_{1g}(1)$, $y$-polarized anti-bonding $B_{2u}(1)$, and $x$-polarized bonding $B_{3u}(1)$. Modes with index $g$ have even (gerade) parity under inversion; modes with index $u$ have odd (ungerade) parity. The representation of the in-plane quadrupole moment is $A_g\oplus B_{1g}$. Although it differs from the dipole representation, the quadrupole induces the same set of irreducible representations in the dimer, see Table~\ref{TAB:plasmon_irreps}. The projected eigenmodes are shown in Fig.~\ref{FIG:dimer_em}. Modes within one column belong to the same representation and have identical selection rules in response to any perturbation. We will discuss the signatures of the dipole- and quadrupole-derived modes in the optical spectra further below.

The hexamer belongs to the $D_{6h}$ point group. This point group is also found in hexagonally packed colloidal layers and crystals.\cite{MuellerFaradayDiscussions2019} The atomic representation of the disc hexamer is $\Gamma_{ar}=A_{1g}\oplus E_{2g}\oplus B_{2u}\oplus E_{1u}$. The in-plane disc dipole belongs to the $E_{1u}$ representation of $D_{6h}$, see Table~\ref{TAB:moments}. The in-plane dipole moment of the disc induces the following representations in the hexamer
\begin{equation}
\begin{split}
\Gamma_{pl}(dip, i)&=E_{1u}\otimes(A_{1g}\oplus E_{2g}\oplus B_{2u}\oplus E_{1u})
\\&=A_{1g}\oplus A_{2g}\oplus 2E_{2g}\oplus B_{1u}\oplus B_{2u}\oplus 2E_{1u}.
\end{split}
\end{equation}
 We  project the dipole-induced, in-plane eigenmodes as shown in Fig.~\ref{FIG:hexamer_em}(a). The non-degenerate $A$ and $B$ eigenmodes have constant amplitude around the hexagon. In the degenerate $E$ modes the amplitude varies around the circumference:  $E_1$ eigenmodes have two and $E_2$ eigenmodes have four nodes around the hexagon.\cite{ReichBuch, Bozovic1985}  The quadrupole mode of the disc induces a second set of hexamer eigenmodes with identical symmetry to the disc dipole. We show the $A$, $B$, and $E_1$ modes that were induced by the quadrupole in Fig.~\ref{FIG:hexamer_em}(b).

 The symmetry-adapted eigenmodes allow comparative predictions on the strength of optical absorption. For example, the $E_{1u}$ eigenmodes will contribute to the absorption of linearly polarized light as we will discuss in detail below. The absorption intensity depends on the number and amplitude of the electromagnetic hotspots. Hotspots are places of very high electric field amplitude that form through near-field coupling from two adjacent discs.\cite{Prodan2003,Forestiere2013} A strong hotspot requires that two adjacent discs face each other with areas of opposite accumulated charge. An inspection of the $E_{1u}$ eigenmodes in Fig.~\ref{FIG:hexamer_em} shows that the number and strength of the hotspots differs from one mode to the other. The $E_{1u}(1)$ mode has four hotspots close to the amplitude maximum in the left eigenvector and two strong hotspots in the right eigenvector. Such a  mode will efficiently absorb and emit far-field radiation; it will also contribute strongly to plasmon-enhanced optical processes such as SERS.\cite{EtchegoinBuch, Langer2019, LeRu2006} The $E_{1u}(2)$ eigenmode, in contrast, has only two hotspots close to the point of vanishing amplitude in one eigenvector and no hotspot between the discs in the other eigenvector. We expect less radiative interaction with far-field photons. The $E_{1u}(3)$ mode forms interparticle hotspots comparable to $E_{1u}(1)$, and we expect strong light absorption by this mode although it was derived from an optically forbidden excitation of the monomer.

\begin{figure}
  \includegraphics[width=8.5cm]{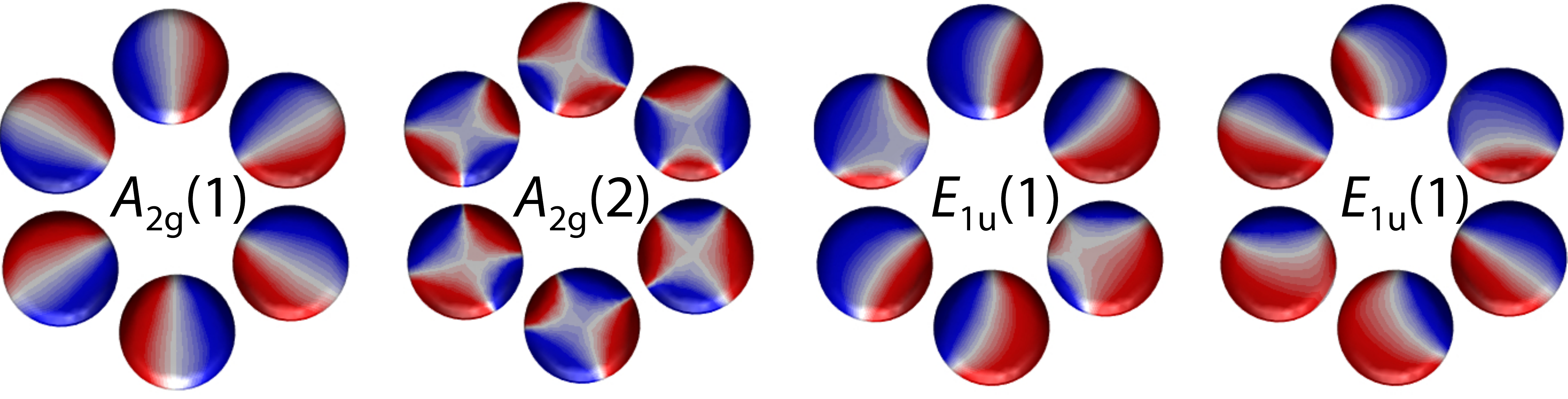}
\caption{$A_{2g}(1), A_{2g}(2)$, and $E_{1u}(1)$ eigenmodes calculated for the hexamer within the boundary elements method. }
\label{FIG:hexamer_bem}
\end{figure}

Eigenmodes that belong to the same irreducible representation of the oligomer are allowed to mix. The real eigenvectors will be a superposition of various monomer excitations.\cite{Pascale2019} The mixing will increase with decreasing gap between the particles, because Coulomb interaction between the monomers alters the charge distribution. Formally, this is described as a contribution by higher-order multipoles of the monomer.\cite{Pascale2019} Although the mixing affects the eigenvectors, the selection rules remain strictly applicable, since the mode symmetry has to be identical. For typical nanooligomers, the calculated eigenmodes remain predominantly dipole-like, quadrupole-like and so forth. To demonstrate this we show the calculated $A_{2g}$ modes and the $E_{1u}(1)$ mode of the hexamer in Fig.~\ref{FIG:hexamer_bem}.  The $A_{2g}(1)$ eigenstate is energetically well separated from the other $A_{2g}$ modes of the hexamer. The calculated eigenvector is essentially identical to the symmetry-adapted mode. The $E_{1u }(1)$ eigenmode, in contrast, has a contribution from a quadrupole-induced excitation as is most visible in the left eigenvector.

We also projected the eigenmodes for a regular trimer, a tetramer, and a pentamer; the symmetry-adapted eigenmodes are shown in Supplementary Figs. S1-S3. The modes show similar features as discussed for the dimer and hexamer above. In particular, there are always dipole- and quadrupole-induced eigenmodes that belong to the same irreducible representation of the oligomer.

\section{Optical selection rules: Linear polarization and cylindrical vector beams}
\label{sec:AllResults}

Optical selection rules predict whether a given transition is allowed by considering the symmetry of the system and the incoming photon. If the system size is much smaller than the wavelength of the light, transitions that transform like the vector representation are allowed by symmetry  to interact with the dipole moment of the electromagnetic field.\cite{InuiBook,WilsonBook, CardonaBuch} Plasmonic and dielectric oligomers, however, are comparable in size to the focus of an incoming light beam. This activates a new set of optical transitions if structured light is used for excitation. In this section we will consider linear polarization and cylindrical vector beams.

Optical absorption from the ground state excites eigenmodes that transform like the incoming photon.\cite{InuiBook, WilsonBook} The incoming light acts as a perturbation with symmetry $\Gamma_\mathcal{H}$ on an initial plasmonic state $\Psi_{pl}^i$ with symmetry $\Gamma_i$. The final state is denoted by $\Psi_{pl}^f$ and symmetry $\Gamma_f$. This transition will be allowed if the direct product\cite{InuiBook}
\begin{equation}
\Gamma_f\otimes \Gamma_\mathcal{H}\otimes\Gamma_i\supset\Gamma_1,
\label{EQ:GT}
\end{equation}
where $\Gamma_1$ is the totally symmetric representation of the point group. We assume that the initial state is the oligomer ground state belonging to $\Gamma_1$. Then Eq.~\eqref{EQ:GT} is equivalent to requiring $\Gamma_f\subset\Gamma_\mathcal{H}$. That means that the representation of the light needs to contain the irreducible representation of the oligomer eigenstate.

The representation of the optical dipole interaction Hamiltonian is given by the vector representation in case of linearly polarized light. The polarization patterns of radial and azimuthal polarization are shown in Fig.~\ref{FIG:dimer_spectra}(a) further below. We find their representations by inspecting the transformation of the polarization patterns under the symmetry operations of the oligomer point groups.\cite{InuiBook, CardonaBuch} Table~\ref{TAB:sel} lists the selection rules we obtained. Linearly polarized light excites dipole-type eigenmodes of the oligomer.\cite{Chuntonov2011,Kerber2017,Jorio2017}  Cylindrical vector beams excite modes with vanishing dipole moment that are normally considered dark.\cite{Hentschel2013,Yanai2014,Parramon2012} Radially polarized beams interact with excitations that belong to the totally symmetric representation. Light with azimuthal polarization will be absorbed by states that transform like the rotation around the $z$ axis within the point group of the oligomer. We expect the optical absorption spectra to change drastically when varying polarization.

\begin{table}
\caption{Optical selection rules for linear, radial, and azimuthal polarization. The light propagation direction is along $z$. $\lambda\rightarrow\infty$ implies the quasi-static approximation where the electric field is considered to be translationally invariant along the propagation direction.}
\label{TAB:sel}
\begin{tabular}{lccc}\hline\hline
&linear&radial&azimuthal\\\hline
\multicolumn{4}{l}{$\lambda\rightarrow\infty$}\\\hline
$D_{2h}$&$B_{3u}(x), B_{2u}(y)$&$A_{1g}$&$B_{1g}$\\
$D_{3h}$&$E'(x,y)$&$A'_{1}$&$A'_{2}$\\
$D_{4h}$&$E_u(x,y)$&$A_{1g}$&$A_{2g}$\\
$D_{5h}$&$E_1'(x,y)$&$A'_{1}$&$A'_{2}$\\
$D_{6h}$&$E_{1u}(x,y)$&$A_{1g}$&$A_{2g}$\\
$C_{2v}$&$B_1(x), B_2(y)$&$A_{1}$&$A_{2}$\\
$D_{\infty h}$&$E_{1u}(x,y)$&$A_{1g}$&$A_{2u}$\\
$C_{\infty v}$&$E_1(x,y)$&$A_{1}$&$A_{2}$\\\hline\hline
\end{tabular}
\end{table}

\subsection{Light scattering and Fano resonances}
Elastic or Rayleigh scattering of light is a prime characterization tool for nanoplasmonic and nanophotonic oligomers. Resonant Rayleigh scattering also known as dark field spectroscopy detects excitations with very high sensitivity.\cite{Knight2010, Crut2014, Wang2019, Kuznetsovaag2016} The symmetry-imposed selection rules of Rayleigh scattering allow any eigenstate as intermediate scattering state, but resonances occur only if the energy and the symmetry of the excited state match the incoming photon. 

In Rayleigh scattering, an incoming photon with symmetry $\Gamma_\mathcal{H}$ excites the system into the intermediate state $\Gamma_n$. The light is immediately re-emitted into the scattered photon with $\Gamma_\mathcal{H}$ symmetry. Since the square of any representation contains the totally symmetric representation, Rayleigh scattering is allowed for any intermediate state, i.e., 
\begin{equation}
\Gamma_f\otimes\Gamma_\mathcal{H}\otimes\Gamma_\mathcal{H}\otimes\Gamma_i\supset\Gamma_1
\label{EQ:Rayleigh}
\end{equation}
is true irrespective of the intermediate state ($\Gamma_f=\Gamma_i$ is the ground state). Resonant Rayleigh scattering, in addition, requires the intermediate excited state to coincide with an eigenstate of the plasmonic system. This will occur if the symmetry of the intermediate state $\Gamma_n$ is contained in $\Gamma_\mathcal{H}$ and the photon energy matches the eigenenergy of $\Gamma_n$. Resonances increase the cross section for light scattering by several orders of magnitude making resonant scattering dominant in the Rayleigh spectra.\cite{Knight2010, Crut2014, Wang2019}

Dark field spectra of plasmonic oligomers often show Fano resonances that arise from the superposition of scattering channels.\cite{Lukyanchuk2010,Miroshnichenko2010,Francescato2012,Forestiere2013,Hopkins2013} Fano resonances may result in anti-resonances in the spectra, i.e., a broad scattering peak with a strong dip at the energy of a second excitation.\cite{Gallinet2013} Fano resonances occur if identical initial and final state are connected by more than one scattering pathway. For resonant Rayleigh scattering this means that two plasmonic excitations of symmetry $\Gamma_n$ contribute to the resonance, because they overlap in excitation energy. A typical case is one superradiant plasmonic mode with a large full width at half maximum (FWHM) that overlaps with a narrow mode with smaller oscillator strength and line width.\cite{Hao2009,Hentschel2011,Forestiere2013,Hopkins2013} We will show that Fano resonances occur for linear polarization in all plasmonic oligomers with three-fold or higher principle axis of rotation.\cite{Forestiere2013,Hopkins2013} An understanding of the symmetry properties of plasmonic eigenmodes allows to tailor Fano resonances by manipulating the oligomer geometry.

\subsection{Spectra of plasmonic oligomers}\label{SEC:results}

\begin{figure}
  \includegraphics[width=8.5cm]{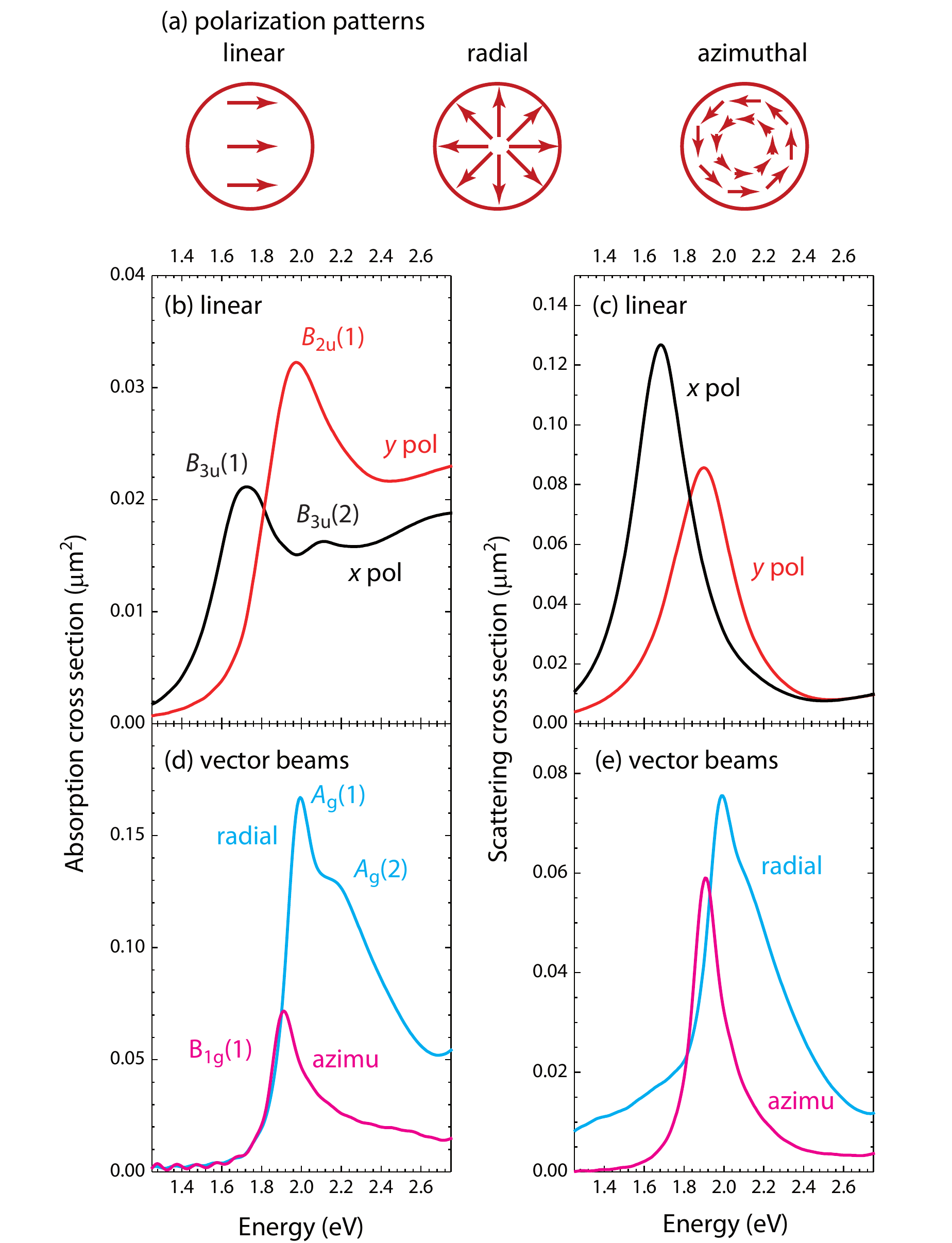}
\caption{(a) Direction of the electric field across the beam focus for linear, radial, and azimuthal polarization. Radial and azimuthal polarization contain a vortex at the beam center. (b)-(e) Optical spectra of a disc dimer ($d=100\,$nm, $h=40\,$nm, $g=20\,$nm, and $\varepsilon=1.65$). (b) Absorption and (c) scattering cross section for linearly polarized light. Black line: $x$ polarization, red line: $y$ polarization. (d) Absorption and (e) scattering cross section for excitation by cylindrical vector beams. Cyan line: radial polarization, magenta: azimuthal polarization. The labels indicate the eigenmode assignment, see Fig.~\ref{FIG:dimer_em}.}
\label{FIG:dimer_spectra}
\end{figure}

In this section we exemplary present optical absorption and elastic light scattering for  plasmonic oligomers. We simulated spectra for linear polarization and cylindrical vector beams, see Fig.~\ref{FIG:dimer_spectra}(a). We will relate the peaks to the projected eigenmodes of the oligomers. Cylindrical vector beams excite dark, non-degenerate plasmon modes that have  narrower line width than bright eigenmodes. For the higher-order oligomers the azimuthal mode is the plasmon of lowest energy. Since all irreducible representations contain more than one plasmon eigenstate, Fano resonances are predicted in the scattering spectra. They are particularly pronounced in linear excitation, because of the large FWHM of the bright plasmon modes.

We first consider a nanodimer with $D_{2h}$ symmetry.
The $x$ and $y$ polarized absorption spectra, Fig.~\ref{FIG:dimer_spectra}(b), each show a dominant peak that arises from the dipole-induced $B_{2u}(1)$ and $B_{3u}(1)$ modes. The spectra contain additional weaker peaks; most pronounced is the quadrupole-induced $B_{3u}(2)$ mode in $x$ polarization. The quadrupole is optically forbidden in the monomer, but gets activated in the dimer by combining two quadrupoles out of phase, which leads to a hotspot in the dimer void. Effectively, the hotspot provides a way to interact with far-field radiation. With decreasing gap size (stronger hotspot) the $B_{3u}(2)$ peak becomes more and more pronounced in the absorption spectrum, see Suppl.\ Fig.\ S5. The $B_{3u}(1)$ mode continues to have the highest integrated intensity, but the peak height is small because of the large FWHM due to the radiative decay of this superradiant mode. The two $B_{3u}$ modes overlap in excitation energy. They will interfere in light scattering creating the characteristic Fano dip close to 2\,eV in Suppl.\ Fig.~S4. 

When exciting the dimer with radially and azimuthally polarized light, the optical spectra change drastically in the energies of the peaks, their FWHM, and their intensity. Cylindrical vector beams excite gerade representations that are optically inactive in the quasi-static dipole approximation. The activation occurs because the  dimer is quite large (220\,nm) compared to the wavelength of light (900-450\,nm in Fig.~\ref{FIG:dimer_spectra}). The right and left disc interact with electromagnetic fields of antiparallel polarization. 
Interestingly, the absorption cross section of the $A_{1g}(1)$ and $B_{1g}(1)$ plasmons in Fig.~\ref{FIG:dimer_spectra}(d) is by a factor of two to five higher than for linearly polarized light in Fig.~\ref{FIG:dimer_spectra}(b). Although cylindrical vector beams get absorbed by the $A_{1g}$ and $B_{1g}$ plasmons, these states do not radiate efficiently into the far field. Light scattering represents a combined excitation and radiation event. The scattering cross section of the vector beams, Fig.~\ref{FIG:dimer_spectra}(e), is much weaker than the absorption cross section, Fig.~\ref{FIG:dimer_spectra}(d), and scattering by linearly polarized light, Fig.~\ref{FIG:dimer_spectra}(c). Strong absorption combined with weak scattering (or radiation) is interesting for several reasons. The small probability for radiation into the far field increases the radiative lifetime of the plasmon eigenmode and reduces its broadening as we discuss in the next section. Also, low scattering and strong absorption effectively cloaks strong scatterers like plasmonic nanostructures.\cite{AndreaAlu2009}

\begin{figure}
  \includegraphics[width=8.5cm]{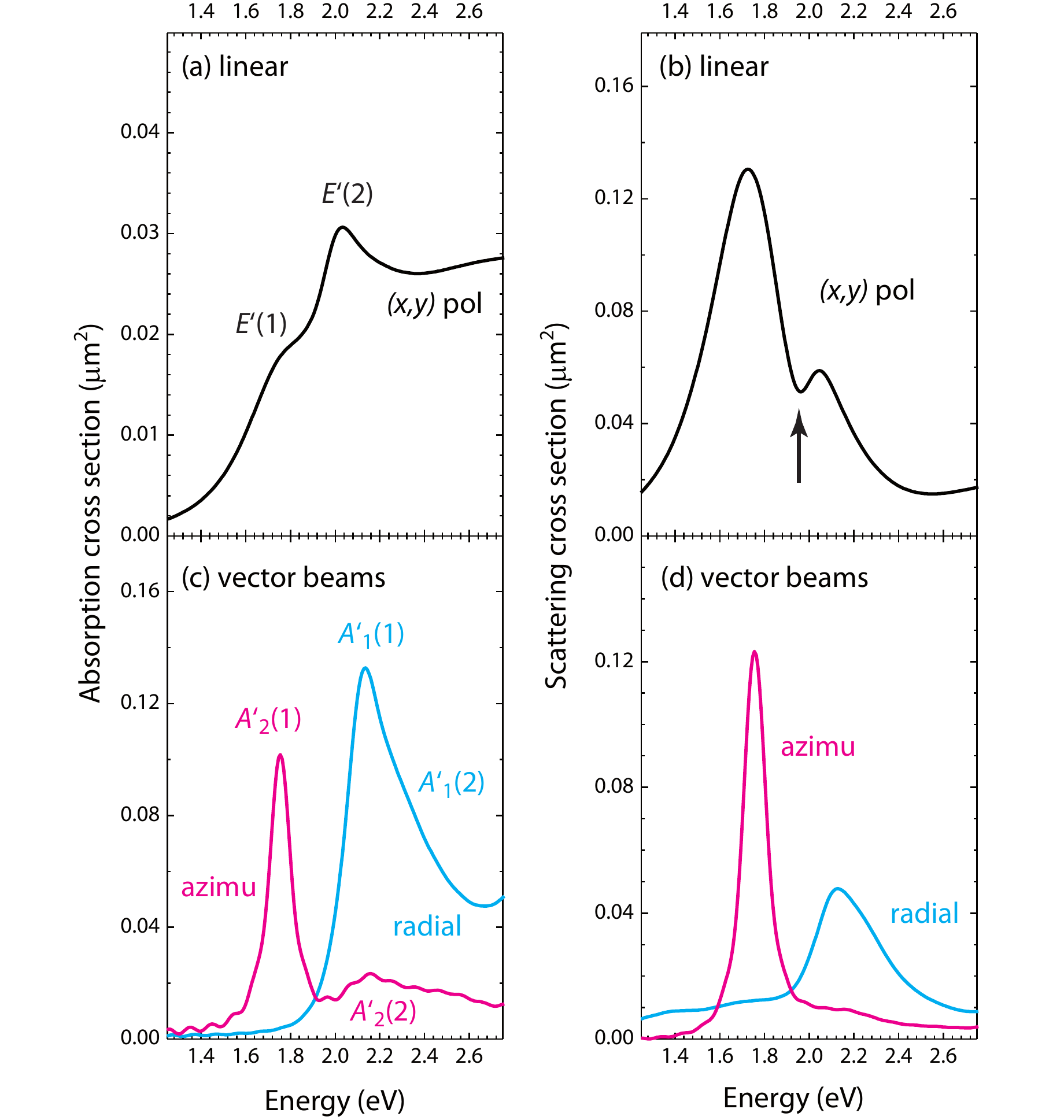}
\caption{Optical spectra of a disc trimer ($d=100\,$nm, $h=40\,$nm, $g=20\,$nm, and $\varepsilon=1.65$). (a) Absorption and (b) scattering for linearly polarized light. The arrow indicates a dip in the scattering cross section that comes from the interference of the $E'(1)$ and $E'(2)$ modes. (c) Absorption and (d) scattering for excitation by cylindrical vector beams. Cyan line: radial polarization, magenta: azimuthal polarization. The labels indicate the eigenmode assignment, see Suppl. Fig.~S1. }
\label{FIG:trimer_spectra}
\end{figure}

A trimer belongs to the $D_{3h}$ point group, Fig.~\ref{FIG:oligomers}.  Figure~\ref{FIG:trimer_spectra} shows the calculated absorption and scattering spectra under linear, radial, and azimuthal polarization. In $D_{3h}$ and all groups with a higher order of principle axis of rotation, the $x$ and $y$ direction are degenerate. Therefore, in-plane linearly polarized light ($E'$ representation) will yield the absorption spectrum in Fig.~\ref{FIG:trimer_spectra}(a) and the scattering spectrum in Fig.~\ref{FIG:trimer_spectra}(b) irrespective of the polarization direction within the plane. Table~\ref{TAB:plasmon_irreps} lists two dipole-induced eigenmodes belonging to the $E'$ representation. Indeed, the absorption spectrum shows two peaks $E'(1)$ and $E'(2)$ that result in a Fano feature in light scattering [arrow in Fig.~\ref{FIG:trimer_spectra}(b)]. In contrast to the dimer where the Fano feature arose from interference between a dipole- and a quadrupole-derived eigenmode, the two peaks in the trimer are induced by monomer dipoles. Absorption and scattering by the quadrupole-induced $E'(3)$ mode, Suppl. Fig.~S1, is too weak to be identified in the spectra. A similar mode will appear more prominently in the higher-order oligomers.

Cylindrical vector beams yield much narrower plasmon resonances than linearly polarized light. Especially, the azimuthal spectrum is remarkable for its small FWHM $\gamma = 107\,$meV, Fig.~\ref{FIG:trimer_spectra}(c). Bulk damping at the $A_2'(1)$ plasmon energy (1.75\,eV) is $\approx70\,$meV,\cite{JohnsonChristy1972,Gallinet2013} so that the contribution from radiative damping appears to be very small. For comparison, the $E'(1)$ mode at almost the same energy (1.74\,eV) has $\gamma=380\,$meV. The simulation highlights the advantage of the dipole-forbidden plasmon modes that cannot decay easily by coupling to the photonic far field.

\begin{figure}
  \includegraphics[width=8.5cm]{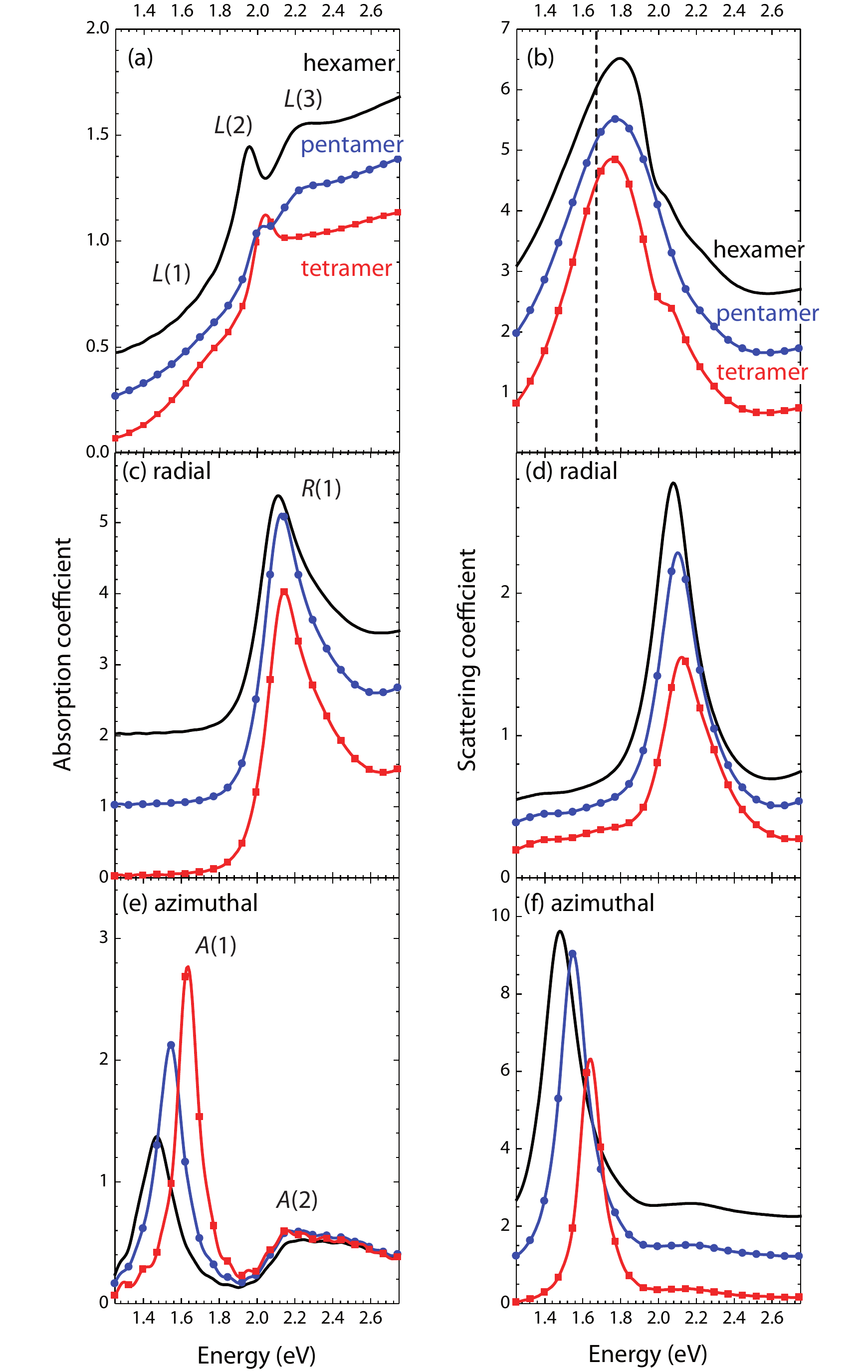}
\caption{Absorption and scattering coefficient for higher-order oligomers: tetramer (red, square), pentamer (blue, dots), and hexamer (black, line). (a) Absorption and (b) scattering coefficient for linear in-plane polarization. The labels stand for the following modes: tetramer -- $L(1)=E_u(1)$, $L(2)=E_u(2),$ and $L(3)=E_u(3)$, pentamer -- $L(1)=E_1'(1)$, $L(2)=E_1'(2)$, and $L(3)=E_1'(3)$, and hexamer -- $L(1)=E_{1u}(1)$, $L(2)=E_{1u}(2)$, and $L(3)=E_{1u}(3)$. (c) Absorption and (d) scattering coefficient for radial in-plane polarization. Labels: tetramer and hexamer -- $R(1)=A_{1g}(1)$ and pentamer $R(1)=A_1'(1)$. (e) Absorption and (f) scattering coefficient for azimuthal in-plane polarization. Labels: tetramer and hexamer -- $A(1)=A_{2g}(1)$ and $A(2)=A_{2g}(2)$, pentamer -- $A(1)=A_2'(1)$ and $A(2)=A_2'(2)$. Except for panel (e), the spectra were shifted vertically for clarity.}
\label{FIG:oligomer_linear}
\end{figure}

The optical properties of higher-order oligomers -- tetramer, pentamer, and hexamer -- evolve incrementally, see Fig.~\ref{FIG:oligomer_linear}. All have degenerate in-plane polarized $(x,y)$ representations.  The disc dipole and quadrupole induce a total of four linearly polarized in-plane oligomer eigenstates. The linearly polarized absorption spectra contain one pronounced peak $L(2)$ and two weaker features at lower $L(1)$ and higher $L(3)$ energy as shown in Fig.~\ref{FIG:oligomer_linear}. They arise from the two dipole-induced and the lowest-energy quadrupole-induced eigenmodes. Quite remarkably, the broadening of some of the peaks is so strong that the most prominent dipole modes are hardly visible in the absorption spectra. For example, $L(1)=E_{1u}(1)$ in the hexamer, Fig.~\ref{FIG:hexamer_em}. This mode had the strongest hotspots of the $E_{1u}$ states resulting in strong far-field coupling and a smeared-out peak with $\gamma=570\,$meV. The $E_{1u}(1)$ mode dominates, however, the scattering spectrum in Fig.~\ref{FIG:oligomer_linear}(b) where the other two $E_{1u}$ modes appear as kinks and dips. The scattering spectra of the higher-order oligomers are remarkably asymmetric, which is a result of interferences between the resonantly scattering modes. The energy of the maximum intensity is higher than the eigenenergy of the L(1) state as shown for the hexamer by the vertical line in Fig.~\ref{FIG:oligomer_linear}(b). This shift needs to be kept in mind when extracting plasmon energies from dark-field spectra.

The higher-order oligomers absorb radially and azimuthally polarized light, Fig.~\ref{FIG:oligomer_linear}(c)-(f). The absorption cross section is very high; it exceeds their geometrical cross section by up to a factor of four. Azimuthally polarized light is also strongly scattered. The peak scattering intensity of $A(1)$ is higher than the linearly polarized $L(1)$ peak. The $A(1)$ peak position shifts to smaller energies with increasing oligomer order making it the state with the smallest energy for $n \geq 3$. At the same time, its scattering intensity and FWHM increase. This behavior reflects the increase in the number of hotspots that simultaneously reduces the plasmon energy in the bonding configuration and increases the coupling to far field radiation. The radially polarized $R(1)$ mode has almost constant eigenenergy and a much smaller increase in the ratio between light scattering and absorption. The totally symmetric $R(1)$ eigenmodes produce no strong hotspots and the number of monomers in the oligomer is less important.

\subsection{Radiative and non-radiative decay}
\label{sec:PlasmonDecay}

\begin{figure}
  \includegraphics[width=8.5cm]{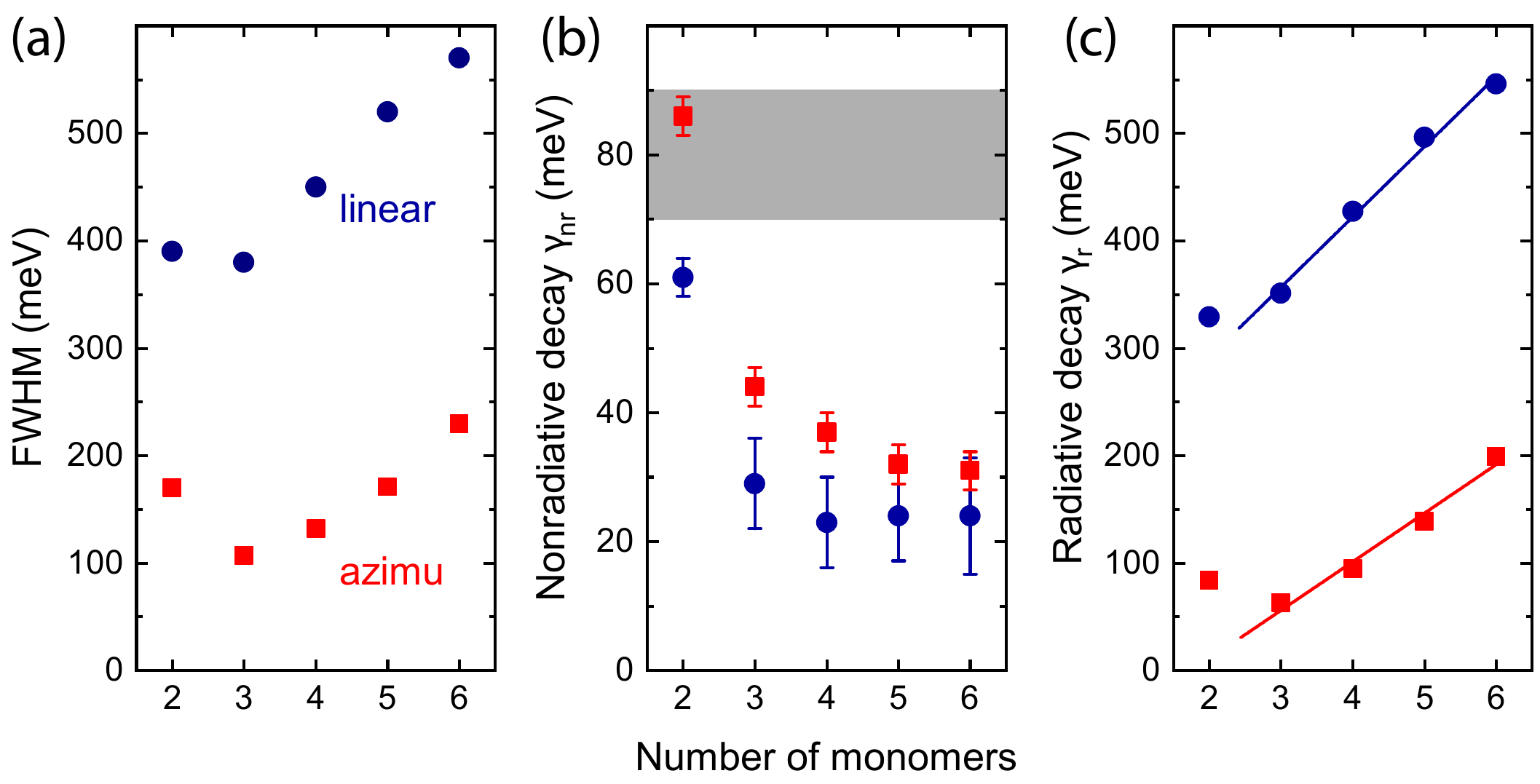}
\caption{(a) FWHM $\gamma$, (b) non-radiative $\gamma_{nr}$, and (c) radiative $\gamma_r$ damping for oligomers of order $n=2-6$. Blue dots are for the lowest-energy dipole allowed transition [dimer $B_{3u}(1)$, trimer $E'(1)$, tetramer $E_u(1)$, pentamer $E_1'(1)$, and hexamer $E_{1u}(1)$] and red squares for the lowest-energy transition for azimuthally polarized light [dimer $B_{1g}(1)$, trimer $A_2'(1)$, tetramer $A_{2g}(1)$, pentamer $A_2'(1)$, and hexamer $A_{2g}(1)$]. The gray area in (b) marks the range of non-radiative damping in bulk gold for all simulated plasmon energies. The lines in panel (c) are a guide to the eye. The error is within the size of the symbols except for panel (b) where error bars are shown.}
\label{FIG:rad_nonrad}
\end{figure}

Many applications of plasmonic and nanophotonic systems require engineering the radiative and non-radiative decay. Plasmon-enhanced spectroscopy, for example, relies on increasing the radiative damping of a nearby dipole via radiating plasmons.\cite{LeRu2006, Li2017, Langer2019} For hot-electron generation, on the other hand, the non-radiative relaxation should be maximized at the expense of radiative decay.\cite{Hartland2017, MuellerFaradayDiscussions2019, Hoeing2020} We will now analyse the radiative $\gamma_r$ and non-radiative $\gamma_{nr}$ decay in the plasmonic oligomers using the simulated spectra.

The ratio of the scattering $\sigma_{sca}$ and absorption $\sigma_{abs}$ cross section is related to the two relaxation channels, as $\gamma_r/\gamma_{nr} = \sigma_{sca}/\sigma_{abs}$.\cite{Liu2009} Using the fact that the total decay (FWHM) is $\gamma=\gamma_r+\gamma_{nr}$, we obtain 
\begin{equation}
    \gamma_{nr}=\gamma(1+\sigma_{sca}/\sigma_{abs})^{-1}.
\end{equation}
As examples we analysed the linearly $L(1)$ and azimuthally $A(1)$ polarized modes with the lowest energy, see caption of Fig.~\ref{FIG:rad_nonrad}. For both plasmons the FWHM increases strongly with increasing order of the oligomer, Fig.~\ref{FIG:rad_nonrad}. This increase in $\gamma$ is entirely caused by the rising radiative decay. The non-radiative decay $\gamma_{nr}$ drops from $80$\,meV close to the bulk value in the dimer [see gray area in Fig.~\ref{FIG:rad_nonrad}(b)] to $20-30\,$meV in the higher-order oligomers.  This corresponds to $30-40\%$ of the bulk damping rate at the energies of the plasmon modes.\cite{Wang2006} For the hexamer only $5-10\%$ of the FWHM is caused by non-radiative decay channels. The reason is that a large fraction of the plasmon mode energy is stored in the oligomer hotspots.\cite{Gallinet2013} This reduces the overlap with the metal electrons and thus non-radiative decay, but increases the radiative damping. We found that $\sim90\%$ of the $A(1)$ mode volume is outside the metal in the hexamer, see Suppl.\ Sect.\ S3, in excellent agreement with its contribution to $\gamma$. The small contribution of $\gamma_{nr}$ is quite remarkable; it implies that a higher quality of the plasmonic material -- e.g. single crystals of a metal -- will hardly affect losses in oligomers with $n>3$. The radiative decay $\gamma_r$ increases linearly with oligomer order, see lines in Fig.~\ref{FIG:rad_nonrad}(c). This is equally true for the linearly polarized (bright) and the azimuthal (dark) mode, although radiative damping of the azimuthal modes remains smaller than for the linearly polarized excitations. Nevertheless, the notion of a ``dark" or ``forbidden" mode is clearly no longer justified for reasonably large plasmonic oligomers.

\section{Light with orbital angular momentum}\label{SEC:OAM}

\begin{figure}
  \includegraphics[width=8.5cm]{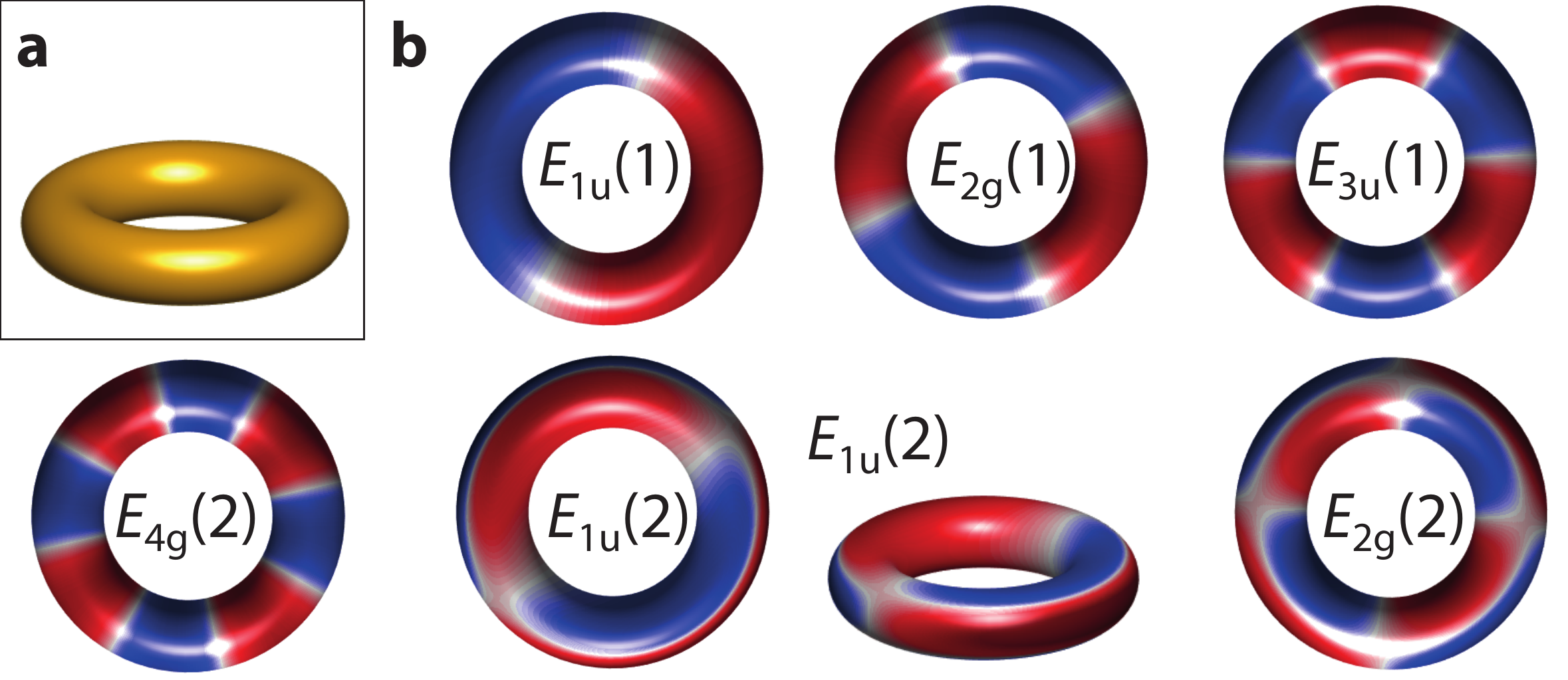}
\caption{(a) Nanoscale ring or torus. (b) Plasmon eigenmodes of the ring. Blue (red) areas stand for positive (negative) surface charge density.}
\label{FIG:ring}
\end{figure}

Another class of structured light are beams that carry orbital angular momentum (OAM).\cite{Allen1992,Yao2011,OAMCollection2017} Since angular momentum is a preserved quantity, we expect novel selection rules for OAM beams. This was elegantly confirmed in recent experiments that examined quadrupole excitations in ions, i.e., transitions that get induced by the quadrupole moment of the electromagnetic field.\cite{Schmiegelow2016,Afanasev2018} Different transitions were excited by OAM beams when varying the magnitude and sign of the orbital angular momentum. For the orbital  momentum to have an effect on dipole excitations, however, the size of the absorbing structure needs to be comparable to the focused beam. Then, the angular momentum is conserved for the entire structure during light absorption and scattering,\cite{Kerber2017,Kerber2018,Machado2018,Konzelmann2019} which excites plasmon eigenmodes with an angular momentum that matches the momentum of the incoming beam.

\begin{table*}
\begin{tabular}{cccccc}\hline\hline
$l_z$&radial&azimu.&left circ.&right circ.&linear\\\hline
\multicolumn{6}{l}{ring, $D_{\infty h}$}\\\hline
0&$A_{1g}$&$A_{2g}$&$E_{1u}$&$E_{1u}$&$E_{1u}$\\
1&$E_{1u}$&$E_{1u}$&$E_{2g}$&$A_{1g}\oplus A_{2g}$&$A_{1g}\oplus A_{2g}\oplus E_{2g}$\\
2&$E_{2g}$&$E_{2g}$&$E_{3u}$&$E_{1u}$&$E_{1u}\oplus E_{3u}$\\
3&$E_{3u}$&$E_{3u}$&$E_{4g}$&$E_{2g}$&$E_{2g}\oplus E_{4g}$\\
even, $\ge 2$&$E_{l_zg}$&$E_{l_zg}$&$E_{(l_z+1)u}$&$E_{(l_z-1)u}$&$E_{(l_z+1)u}\oplus E_{(l_z-1)u}$\\
odd, $\ge 3$&$E_{l_zu}$&$E_{l_zu}$&$E_{(l_z+1)g}$&$E_{(l_z-1)g}$&$E_{(l_z+1)g}\oplus E_{(l_z-1)g}$\\\hline
\multicolumn{6}{l}{hexamer, $D_{6h}$}\\\hline
0&$A_{1g}$&$A_{2g}$&$E_{1u}$&$E_{1u}$&$E_{1u}$\\
1&$E_{1u}$&$E_{1u}$&$E_{2g}$&$A_{1g}\oplus A_{2g}$&$A_{1g}\oplus A_{2g}\oplus E_{2g}$\\
2&$E_{2g}$&$E_{2g}$&$B_{1u}\oplus B_{2u}$&$E_{1u}$&$B_{1u}\oplus B_{2u}\oplus E_{1u}$\\
3&$B_{1u}$&$B_{2u}$&$E_{2g}$&$E_{2g}$&$E_{2g}$\\
4&$E_{2g}$&$E_{2g}$&$B_{1u}\oplus B_{2u}$&$E_{1u}$&$B_{1u}\oplus B_{2u}\oplus E_{1u}$\\\hline\hline
\end{tabular}
\caption{Selection rules for OAM beams within the $D_{\infty h}$ and $D_{6h}$ point groups. The selection rules for $l_z<0$ are obtained by flipping the sign of $l_z$ and $m_p$ simultaneously.}
\label{TAB:OAM}
\end{table*}

To study the optical selection rules for OAMs within group theory, we first consider a ring with nanoscale dimensions, Fig.~\ref{FIG:ring}a. The structure belongs to the $D_{\infty h}$ point group. The eigenstates of the ring are standing waves around the circumference, Fig.~\ref{FIG:ring}.\cite{Aizpurua2003} The wavelength of these excitations $\lambda$ inside the material is given by $\pi d_r = m \lambda$, where $d_r$ is the diameter of the ring. The integer $m$ can be identified with the $z$ component of the angular momentum with respect to the principle axis; it is a conserved quantity.\cite{InuiBook,Damnjanovic1999,ReichBuch,Bozovic1985} To specify an eigenstate of the ring we need a set of three quantum numbers, see Methods:\cite{DamnjanovicBuch,ReichBuch,Bozovic1985} The $z$ component of the angular momentum $m$ as introduced above, the parity of the horizontal mirror plane $\sigma_h$, and the parity of the mirror planes $\sigma_v$ that contain the $z$ axis. To obtain the $m$ quantum number for the state excited by an OAM, we have to add the orbital angular momentum $l_z$ of the beam and the spin angular momentum $m_p$ related to polarization, $m=l_z + m_p$. 
Radial polarization has $m_p=0$, $\sigma_h=+1$, and $\sigma_v=+1$; azimuthal polarization $m_p=0, \sigma_h=+1$, and $\sigma_v=-1$. For left (right) handed circular polarization $m_p=+1$ ($-1$), $\sigma_h=+1$, and $\sigma_v=\pm1$ is not defined. This means that both the representation for $\sigma_v=+1$ and $\sigma_v=-1$ will contribute for circular polarization. Finally, linear polarization is the superposition of left- and right-handed circularly polarized light. Taken together, we find the selection rules listed in Table~\ref{TAB:OAM} for the ring. They apply to nanostructures with full rotational symmetry around the propagation direction of the OAM ($z$ axis). Ordinary, linearly polarized light ($l_z=0$) excites the $E_{1u}$ modes in Fig.~\ref{FIG:ring}b; a beam with $l_z=+3$ will excite the $E_{3u}$ mode if it is radially or azimuthally polarized, the $E_{4g}$ mode for left-handed, and the $E_{2g}$ modes for right-handed circular polarization, and $E_{2g}$ and $E_{4g}$ modes for linear polarization.  

We now proceed from the ring structure to nanoscale oligomers. The rotational symmetry of the oligomer is described by the principle axis of rotation $C_n$ with order $n$. Because only rotations by certain angles preserve the symmetry of the oligomer, $m$ can only take on integer values with $|m|\le n/2$.\cite{DamnjanovicBuch,Damnjanovic1999,ReichBuch,Bozovic1985} Higher absolute values of $m$ are brought back into the allowed range with an Umklapp rule, where $m$ is replaced by $m'=n-m$.\cite{DamnjanovicBuch,Bozovic1985} We apply the Umklapp rule and combine it with the parity selection rules of the polarization patterns into the OAM selection rules of a hexamer, Table~\ref{TAB:OAM}. We find that through the right combination of OAM and polarization, all eigenmodes of the hexamer become accessible to optical spectroscopy. The selection rules excellently explain the simulated scattering of an OAM beam by a hexamer (and other oligomers).\cite{Kerber2017} For examples, the modes excited for $l_z=2$ and 4 in Ref.~\citenum{Kerber2017} are identical as expected from the selection rules. The breathing-like $A_{1g}$ eigenmode appears for $m=l_z+m_p=-1+1=0$. The parallel and antiparallel spectra for $|l_z|=3$ are identical and so forth. The calculated eigenmodes in Ref.~\citenum{Kerber2017} likewise agree with the projected eigenstates in Fig.~\ref{FIG:hexamer_em} when replacing the discs by rod monomers. The conservation of the $m$ quantum number during excitation also explains the orbital angular momentum dichroism proposed in Ref.~\citenum{Kerber2018}. Since $m$ and not $l_z$ is the conserved quantity, the excited states change when changing the sign of $l_z$, if the light also carries spin momentum.  

Beams with angular momentum will allow addressing a wide range of optical excitations in nanophotonic systems, see Suppl.~Tables~S2 and S3. The angular momentum provides an additional degree of freedom to tailor the properties and light-matter interaction for plasmonic and dielectric modes. Such excitations will produce near-fields with well-defined angular momentum. In this way, angular momentum may be transferred to much smaller nanostructures via plasmon-mediated excitations.

\section{Retardation of the incoming light}\label{SEC:retardation}

In discussing novel selection rules from variations in the electric field, we have focused so far on the spatial extension of the oligomers compared to the focus of the light beam. Since light is an electromagnetic wave, the electric field also varies along its propagation direction at a given time. The field retardation will excite dipole-forbidden modes of the oligomer, if the extension along the propagation direction becomes a sizable fraction of the light wavelength.\cite{Rivera2016} For example, in oligomers with discs consisting of a vertical metal-insulator-metal stack, antiparallel plasmonic dipoles are excited in the upper and lower metal disc.\cite{Pakizeh2006, Verre2015, Chang2012} Similarly, light propagating normal to a gold nanoparticle bilayer excites plasmons with antiparallel dipole moments in the two layers.\cite{MuellerFaradayDiscussions2019,MuellerACSPhotonics2018} 

To understand the selection rules introduced by field retardation, we consider a dimer of two spherical nanoparticles (point group $D_{\infty h}$) and light propagating along its $z$ axis (that is the $C_\infty$ axis of the dimer). The in-plane dipole moment induces an $E_{1u}$ and an $E_{1g}$ dimer eigenmode, Table.~\ref{TAB:plasmon_irreps}. The $E_{1u}$ mode is optically active within the quasi-static approximation.\cite{InuiBook, WilsonBook} To introduce field retardation we assume that half the wavelength matches the center-to-center distance of the two spheres $\lambda = (d+g)/2$. In this situation the electric field points in opposite direction at the two spheres. Field retardation will affect selection rules if the point group contains the inversion and/or horizontal mirror plane. The parity for these operations changes to $-1$.\cite{MuellerFaradayDiscussions2019} This replaces a given gerade representation by its ungerade counterpart and vice versa. Instead of the $E_{1u}$ mode, the $E_{1g}$ eigenstate is allowed for the retarded field and linearly polarized light. This mode has the two dipoles pointing in opposite direction.\cite{MuellerFaradayDiscussions2019,Bruno2019} In a real experiment, the wavelength will neither be infinite nor match the dimer size and both modes will contribute to the optical spectra. 
Indeed, optical experiments on hexagonal layers of nanoparticles observed absorption by a plasmon mode in the bilayer that was absent from the spectrum of a monolayer.\cite{MuellerFaradayDiscussions2019,MuellerACSPhotonics2018} The mode had parallel dipoles within a layer, but antiparallel dipoles from one layer to the next, which corresponds to the $E_{1g}$ mode of the nanosphere dimer.\cite{MuellerFaradayDiscussions2019} The excitation of the plasmon under normal incidence was due to field retardation and the comparatively large nanoparticle diameters (30-50\,nm) used in the experiment.

 The selection rules for  structured light within the quasi-static limit were given by Tables~\ref{TAB:sel} and \ref{TAB:OAM}, which we now extend to the retarded cases. For cylindrical vector beams  with $\lambda=(d+g)/2$, radially polarized light excites the $A_{1u}$ mode of the dimer and azimuthally polarized light the $A_{2u}$ eigenmodes. Observing modified selection rules due to field retardation for structured light requires nanostructures that are a sizable fraction of the focus in the $(x,y)$ plane as well as the wavelength along $z$. Such systems need a careful design to realize the predicted non-standard excitations.

\section{Multi-photon processes}\label{SEC:multi}

Multi-photon processes occur in nonlinear optics: A material gets excited by absorbing two photons, three incoming photons convert into one photon with three times the frequency, or light gets scattered inelastically in Raman or hyper-Raman processes.\cite{ButcherBook,BoydBook,CardonaBuch} In this section we discuss selection rules of exemplary multi-photon processes when using structured light for excitation.\cite{Bautista2018,Kroychuk2019} We will show how OAMs may be used to induce second-harmonic generation in centrosymmetric oligomers. Such an experiment will verify the transfer of angular momentum between the photons and the oligomer. 

\begin{table}
\begin{tabular}{lccc}\hline\hline
point group&linear&radial&azimuthal\\\hline
\multicolumn{4}{l}{2-photon absorption}\\\hline
$D_{2h}$&$A_{g}$&$A_g$&$A_g$\\
$D_{3h}$&$A_1'\oplus A'_2\oplus E'$&$A_1'$&$A_1'$\\
$D_{4h}$&$A_{1g}\oplus A_{2g}\oplus B_{1g}\oplus B_{2g}$&$A_{1g}$&$A_{1g}$\\
$D_{5h}$&$A_1'\oplus A'_2\oplus E_2'$&$A_1'$&$A_1'$\\
$D_{6h}$&$A_{1g}\oplus A_{2g}\oplus E_{2g}$&$A_{1g}$&$A_{1g}$\\
$D_{\infty h}$&$A_{1g}\oplus A_{2g}\oplus E_{2g}$&$A_{1g}$&$A_{1g}$\\\hline
\multicolumn{4}{l}{3-photon absorption}\\\hline
$D_{2h}$&$B_{2u}\oplus B_{3u}$&$A_g$&$B_{1g}$\\
$D_{3h}$&$A_1'\oplus A'_2\oplus E'$&$A_1'$&$A_2'$\\
$D_{4h}$&$E_u$&$A_{1g}$&$A_{2g}$\\
$D_{5h}$&$E_1'\oplus E_2'$&$A_1'$&$A_2'$\\
$D_{6h}$&$B_{1u}\oplus B_{2u}\oplus E_{1u}$&$A_{1g}$&$A_{2g}$\\
$D_{\infty h}$&$E_{1u}\oplus E_{3u}$&$A_{1g}$&$A_{1g}$\\\hline
    \end{tabular}
    \caption{Selection rules for two- and three-photon absorption for linear polarization and cylindrical vector beams.}
    \label{TAB:multi_photon}
\end{table}

In two-photon absorption two incoming photons excite an eigenstate of the system. The group theory treatment is identical to linear absorption except that we consider two perturbations (=photons) with symmetry $\Gamma_\mathcal{H}$. For simplicity, we assume the two photons to have identical symmetry. Equation~(\ref{EQ:GT})
is rewritten as 
\begin{equation}
    \Gamma_f\otimes \Gamma_\mathcal{H}\otimes 
    \Gamma_\mathcal{H}\otimes\Gamma_i\supset\Gamma_1,\label{EQ:2pt}
\end{equation}
which is equivalent to $\Gamma_f\subset \Gamma_\mathcal{H}\otimes \Gamma_\mathcal{H}$. Equation~\eqref{EQ:2pt} appears at first to be identical to the conditions for Rayleigh scattering in Eq.~\eqref{EQ:Rayleigh}, but in two-photon absorption the initial $\Gamma_i=\Gamma_1$ and final state $\Gamma_f$ differ. Table~\ref{TAB:multi_photon} lists the selection rules of the $D_{n h}$ point groups considered here. Cylindrical vector beams always excite the totally symmetric representation of the oligomers in two-photon absorption. Two linearly polarized photons allow exciting states that are forbidden for a single photon as is standard in this technique. Particularly interesting is the trimer ($D_{3h}$) where the $E_1'$ representation is active in one- and two-photon excitation. This means that the trimer will produce a second-harmonic signal (SHG, second-harmonic generation). In none of the other oligomers the dipole-active states contribute to two-photon absorption, which is the standard requirement for SHG activity. However, the absorption of two photons from a cylindrical vector beam should leads to the emission of radially polarized light, which would be extremely interesting to observe. 

When OAM is added as an additional degree of freedom, SHG under emission of linearly polarized photons may be activated in all oligomers as we show now. For this we have to allow for photons with different OAM. First, we consider two photons $p1$ and $p2$ with $l_z=0$,  linear polarization, and an oligomer belonging to $D_{6h}$. The two photons may combine into a state with $m=\pm2$ ($E_{2g}$ eigenstates) or $m=0$ ($A_{1g}, A_{2g}$). We now change the OAM of $p2$ to $l_z^{p2}=+1$. The total angular momentum may add up to $m=3$ ($E_{3u}$) or $m=\pm1$ ($E_{1u}$). The $E_{1u}$ excitation may decay by emitting a single linearly polarized photon. 
We find that exciting a hexamer with two photons that have $l_z^{p1}=0$ and $l_z^{p2}=+1$ will give rise to a second-harmonic signal. SHG will be at maximum if the sum of the two photon energies matches the one-photon transition of the oligomers. This experiment would be particularly interesting to perform, because it proves the transfer of angular momentum from the photon to the oligomer.

Selection rules for other higher-order processes may be derived in a similar manner by reducing the product of the one-photon representations. In Table~\ref{TAB:multi_photon} we list the selection rules for three-photon absorption in oligomers. The representations that are allowed for the emission/absorption by one photon are also reached through three-photon absorption. This means that all oligomers will produce third-harmonic signal. The selection rules for Raman scattering are identical to two-photon absorption; hyper-Raman scattering obeys the selection rules of three-photon absorption, and so forth. A comprehensive group theoretical treatment of surface-enhanced Raman scattering was published by some of us recently.\cite{Jorio2017}

\section{Conclusion}\label{SEC:conclusion}

We derived the optical selection rules in nanoscale systems excited by linearly polarized and structured light. The nanosystems have extensions that are comparable to the wavelength of light and the focus of a light beam. When excited by structured light, the oligomers experience the varying phase and polarization patterns. We derived the selection rules for absorption and scattering of cylindrical vector beams and light with orbital angular momentum considering the dipole moment of the elctromagnetic field. Structured light allows the excitation of oligomer eigenmodes that are dark/optically forbidden under linear polarization. We discussed the changes in the optical spectra for exemplary nanostructures using FDTD simulations of highly symmetric disc oligomers with $n=2-6$ monomers. The radiative and non-radiative decay rate depends systematically on the mode under study as well as the number of monomers. The non-radiative damping rate falls below the lower bound predicted from the quasi-static approximation,\cite{Wang2006} which needs to be considered when engineering plasmonic structures for plasmon-enhanced spectroscopy and hot-electron applications. Using structured light modifies the selection rules in multi-photon processes. Specifically we showed that SHG gets activated when using two linearly polarized photons that differ in their OAM by one. Structured light will unlock a rich world of optical excitations in nanoscale oligomers. Such structures may excite molecules and nanomaterials via their optical near fields. We envision near-field absorption as a way to channel structured light to materials excitations. This would unlock novel excitations and spectroscopic techniques in a wide range of physical systems.

\begin{acknowledgement}
This work was supported by the European Research Council (ERC) within the project DarkSERS (772108) and the Focus Area NanoScale of Freie Universit\"at Berlin. The simulated absorption and scattering spectra of the oligomers are available at the repository REFUBIUM, under identifier http://dx.doi.org/10.17169/refubium-26906.
\end{acknowledgement}

\begin{suppinfo}
Supporting information is provided as a pdf file: Images of the symmetry-adapted eigenvectors in the $D_{3h}, D_{4h}$, and $D_{5h}$ point groups, simulated extinction spectra, table for the correspondence between quantum numbers and irreducible representations in $D_{nh}$ up to $n=6$,  tables for the selection rules for OAM beams for the oligomers considered in this work and the general $D_{qh}$ point groups, and details of methods. This material is available free of charge via the internet at http://pubs.acs.org.
\end{suppinfo}

\bibliography{References_selection}

\end{document}